\documentclass[conference,final]{IEEEtran}
\IEEEoverridecommandlockouts

\usepackage[utf8]{inputenc}
\usepackage[T1]{fontenc}
\usepackage{microtype,newtxtext,newtxmath} 
\usepackage{mathtools,gensymb,eurosym,url,soul} 
\usepackage{subcaption}
\usepackage{longtable,booktabs,array}
\usepackage{cite}
\usepackage[switch]{lineno}
\usepackage{acronym}
\usepackage{graphicx}
\usepackage{xcolor}
\usepackage{xr} 
\usepackage{hyperref}
\hypersetup{colorlinks=true, breaklinks=true, 
  urlcolor=blue, linkcolor=blue,anchorcolor=blue,citecolor=blue,
  hypertexnames=true, final=true, 
  pdfpagemode = UseNone, 
  pdfauthor = {},
  pdftitle = {},   
  pdfsubject = {},
  pdfkeywords = {}
}

\graphicspath{{img/singles/} {img/} {./}}
\DeclareGraphicsExtensions{.pdf,.png,.jpg,.mps}

\acrodef{ADC}[ADC]{Analog to Digital Converter}
\acrodef{ADEXP}[AdExp-I\&F]{Adaptive-Exponential Integrate and Fire}
\acrodef{ADM}[ADM]{Asynchronous Delta Modulator}
\acrodef{AER}[AER]{Address-Event Representation}
\acrodef{AEX}[AEX]{AER EXtension board}
\acrodef{AE}[AE]{Address-Event}
\acrodef{AFM}[AFM]{Atomic Force Microscope}
\acrodef{AGC}[AGC]{Automatic Gain Control}
\acrodef{AI}[AI]{Artificial Intelligence}
\acrodef{AMDA}[AMDA]{AER Motherboard with D/A converters}
\acrodef{ANN}[ANN]{Artificial Neural Network}
\acrodef{API}[API]{Application Programming Interface}
\acrodef{APMOM}[APMOM]{Alternate Polarity Metal On Metal}
\acrodef{ARM}[ARM]{Advanced RISC Machine}
\acrodef{ASIC}[ASIC]{Application Specific Integrated Circuit}
\acrodef{AdExp}[AdExp-IF]{Adaptive Exponential Integrate-and-Fire}
\acrodef{BCM}[BMC]{Bienenstock-Cooper-Munro}
\acrodef{BD}[BD]{Bundled Data}
\acrodef{BEOL}[BEOL]{Back-end of Line}
\acrodef{BG}[BG]{Bias Generator}
\acrodef{BMI}[BMI]{Brain-Machince Interface}
\acrodef{BTB}[BTB]{band-to-band tunnelling}
\acrodef{CAD}[CAD]{Computer Aided Design}
\acrodef{CAM}[CAM]{Content Addressable Memory}
\acrodef{CAVIAR}[CAVIAR]{Convolution AER Vision Architecture for Real-Time}
\acrodef{CA}[CA]{Cortical Automaton}
\acrodef{CCN}[CCN]{Cooperative and Competitive Network}
\acrodef{CDR}[CDR]{Clock-Data Recovery}
\acrodef{CFC}[CFC]{Current to Frequency Converter}
\acrodef{CHP}[CHP]{Communicating Hardware Processes}
\acrodef{CMIM}[CMIM]{Metal-insulator-metal Capacitor}
\acrodef{CML}[CML]{Current Mode Logic}
\acrodef{CMOL}[CMOL]{Hybrid CMOS nanoelectronic circuits}
\acrodef{CMOS}[CMOS]{Complementary Metal-Oxide-Semiconductor}
\acrodef{CNN}[CCN]{Convolutional Neural Network}
\acrodef{COTS}[COTS]{Commercial Off-The-Shelf}
\acrodef{CPG}[CPG]{Central Pattern Generator}
\acrodef{CPLD}[CPLD]{Complex Programmable Logic Device}
\acrodef{CPU}[CPU]{Central Processing Unit}
\acrodef{CSM}[CSM]{Cortical State Machine}
\acrodef{CSP}[CSP]{Constraint Satisfaction Problem}
\acrodef{CTXCTL}[CTXCTL]{CortexControl}
\acrodef{CV}[CV]{Coefficient of Variation}
\acrodef{DAC}[DAC]{Digital to Analog Converter}
\acrodef{DAS}[DAS]{Dynamic Auditory Sensor}
\acrodef{DAVIS}[DAVIS]{Dynamic and Active Pixel Vision Sensor}
\acrodef{DBN}[DBN]{Deep Belief Network}
\acrodef{DFA}[DFA]{Deterministic Finite Automaton}
\acrodef{DIBL}[DIBL]{drain-induced-barrier-lowering}
\acrodef{DI}[DI]{delay insensitive}
\acrodef{DMA}[DMA]{Direct Memory Access}
\acrodef{DNF}[DNF]{Dynamic Neural Field}
\acrodef{DNN}[DNN]{Deep Neural Network}
\acrodef{DOF}[DOF]{Degrees of Freedom}
\acrodef{DPE}[DPE]{Dynamic Parameter Estimation}
\acrodef{DPI}[DPI]{Differential Pair Integrator}
\acrodef{DRAM}[DRAM]{Dynamic Random Access Memory}
\acrodef{DRRZ}[DR-RZ]{Dual-Rail Return-to-Zero}
\acrodef{DR}[DR]{Dual Rail}
\acrodef{DSP}[DSP]{Digital Signal Processor}
\acrodef{DVS}[DVS]{Dynamic Vision Sensor}
\acrodef{DYNAP}[DYNAP]{Dynamic Neuromorphic Asynchronous Processor}
\acrodef{EBL}[EBL]{Electron Beam Lithography}
\acrodef{EDVAC}[EDVAC]{Electronic Discrete Variable Automatic Computer}
\acrodef{EEG}[EEG]{electroencephalography}
\acrodef{EIN}[EIN]{Excitatory-Inhibitory Network}
\acrodef{EM}[EM]{Expectation Maximization}
\acrodef{EMG}[EMG]{electromygraphy}
\acrodef{EPSC}[EPSC]{Excitatory Post-Synaptic Current}
\acrodef{EPSP}[EPSP]{Excitatory Post-Synaptic Potential}
\acrodef{EZ}[EZ]{Epileptogenic Zone}
\acrodef{FDSOI}[FDSOI]{Fully-Depleted Silicon on Insulator}
\acrodef{FET}[FET]{Field-Effect Transistor}
\acrodef{FFT}[FFT]{Fast Fourier Transform}
\acrodef{FI}[F-I]{Frequency-Current}
\acrodef{FPGA}[FPGA]{Field Programmable Gate Array}
\acrodef{FR}[FR]{Fast Ripple}
\acrodef{FSA}[FSA]{Finite State Automaton}
\acrodef{FSM}[FSM]{Finite State Machine}
\acrodef{GIDL}[GIDL]{gate-induced-drain-leakage}
\acrodef{GOPS}[GOPS]{Giga-Operations per Second}
\acrodef{GPU}[GPU]{Graphical Processing Unit}
\acrodef{GUI}[GUI]{Graphical User Interface}
\acrodef{HAL}[HAL]{Hardware Abstraction Layer}
\acrodef{HFO}[HFO]{High Frequency Oscillation}
\acrodef{HH}[H\&H]{Hodgkin \& Huxley}
\acrodef{HMM}[HMM]{Hidden Markov Model}
\acrodef{HRS}[HRS]{High-Resistive State}
\acrodef{HR}[HR]{Human Readable}
\acrodef{HSE}[HSE]{Handshaking Expansion}
\acrodef{HW}[HW]{Hardware}
\acrodef{ICT}[ICT]{Information and Communication Technology}
\acrodef{IC}[IC]{Integrated Circuit}
\acrodef{iEEG}[iEEG]{intracranial electroencephalography}
\acrodef{IF2DWTA}[IF2DWTA]{Integrate \& Fire 2--Dimensional WTA}
\acrodef{IFSLWTA}[IFSLWTA]{Integrate \& Fire Stop Learning WTA}
\acrodef{IF}[I\&F]{Integrate-and-Fire}
\acrodef{IMU}[IMU]{Inertial Measurement Unit}
\acrodef{INCF}[INCF]{International Neuroinformatics Coordinating Facility}
\acrodef{INI}[INI]{Institute of Neuroinformatics}
\acrodef{IO}[I/O]{Input/Output}
\acrodef{IPSC}[IPSC]{Inhibitory Post-Synaptic Current}
\acrodef{IPSP}[IPSP]{Inhibitory Post-Synaptic Potential}
\acrodef{IP}[IP]{Intellectual Property}
\acrodef{ISI}[ISI]{Inter-Spike Interval}
\acrodef{IoT}[IoT]{Internet of Things}
\acrodef{JFLAP}[JFLAP]{Java - Formal Languages and Automata Package}
\acrodef{LEDR}[LEDR]{Level-Encoded Dual-Rail}
\acrodef{LFP}[LFP]{Local Field Potential}
\acrodef{LLC}[LLC]{Low Leakage Cell}
\acrodef{LNA}[LNA]{Low-Noise Amplifier}
\acrodef{LPF}[LPF]{Low Pass Filter}
\acrodef{LRS}[LRS]{Low-Resistive State}
\acrodef{LSM}[LSM]{Liquid State Machine}
\acrodef{LTD}[LTD]{Long Term Depression}
\acrodef{LTI}[LTI]{Linear Time-Invariant}
\acrodef{LTP}[LTP]{Long Term Potentiation}
\acrodef{LTU}[LTU]{Linear Threshold Unit}
\acrodef{LUT}[LUT]{Look-Up Table}
\acrodef{LVDS}[LVDS]{Low Voltage Differential Signaling}
\acrodef{MCMC}[MCMC]{Markov-Chain Monte Carlo}
\acrodef{MEMS}[MEMS]{Micro Electro Mechanical System}
\acrodef{MFR}[MFR]{Mean Firing Rate}
\acrodef{MIM}[MIM]{Metal Insulator Metal}
\acrodef{MLP}[MLP]{Multilayer Perceptron}
\acrodef{MOSCAP}[MOSCAP]{Metal Oxide Semiconductor Capacitor}
\acrodef{MOSFET}[MOSFET]{Metal Oxide Semiconductor Field-Effect Transistor}
\acrodef{MOS}[MOS]{Metal Oxide Semiconductor}
\acrodef{MRI}[MRI]{Magnetic Resonance Imaging}
\acrodef{NDFSM}[NDFSM]{Non-deterministic Finite State Machine} 
\acrodef{ND}[ND]{Noise-Driven}
\acrodef{NEF}[NEF]{Neural Engineering Framework}
\acrodef{NHML}[NHML]{Neuromorphic Hardware Mark-up Language}
\acrodef{NIL}[NIL]{Nano-Imprint Lithography}
\acrodef{NMDA}[NMDA]{N-Methyl-D-Aspartate}
\acrodef{NME}[NE]{Neuromorphic Engineering}
\acrodef{NN}[NN]{Neural Network}
\acrodef{NRZ}[NRZ]{Non-Return-to-Zero}
\acrodef{NSM}[NSM]{Neural State Machine}
\acrodef{OR}[OR]{Operating Room}
\acrodef{OTA}[OTA]{Operational Transconductance Amplifier}
\acrodef{PCB}[PCB]{Printed Circuit Board}
\acrodef{PCHB}[PCHB]{Pre-Charge Half-Buffer}
\acrodef{PCM}[PCM]{Phase Change Memory}
\acrodef{PE}[PE]{Phase Encoding}
\acrodef{PFA}[PFA]{Probabilistic Finite Automaton}
\acrodef{PFC}[PFC]{prefrontal cortex}
\acrodef{PFM}[PFM]{Pulse Frequency Modulation}
\acrodef{PR}[PR]{Production Rule}
\acrodef{PSC}[PSC]{Post-Synaptic Current}
\acrodef{PSP}[PSP]{Post-Synaptic Potential}
\acrodef{PSTH}[PSTH]{Peri-Stimulus Time Histogram}
\acrodef{QDI}[QDI]{Quasi Delay Insensitive}
\acrodef{RAM}[RAM]{Random Access Memory}
\acrodef{RA}[RA]{Resected Area}
\acrodef{RDF}[RDF]{random dopant fluctuation}
\acrodef{RELU}[ReLu]{Rectified Linear Unit}
\acrodef{RLS}[RLS]{Recursive Least-Squares}
\acrodef{RMSE}[RMSE]{Root Mean Squared-Error}
\acrodef{RMS}[RMS]{Root Mean Squared}
\acrodef{RNN}[RNN]{Recurrent Neural Networks}
\acrodef{ROLLS}[ROLLS]{Reconfigurable On-Line Learning Spiking}
\acrodef{RRAM}[R-RAM]{Resistive Random Access Memory}
\acrodef{R}[R]{Ripples}
\acrodef{SAC}[SAC]{Selective Attention Chip}
\acrodef{SAT}[SAT]{Boolean Satisfiability Problem}
\acrodef{SCX}[SCX]{Silicon CorteX}
\acrodef{SD}[SD]{Signal-Driven}
\acrodef{SEM}[SEM]{Spike-based Expectation Maximization}
\acrodef{SLAM}[SLAM]{Simultaneous Localization and Mapping}
\acrodef{SNN}[SNN]{Spiking Neural Network}
\acrodef{SNR}[SNR]{Signal to Noise Ratio}
\acrodef{SOC}[SOC]{System-On-Chip}
\acrodef{SOI}[SOI]{Silicon on Insulator}
\acrodef{SOZ}[SOZ]{Seizure Onset Zone}
\acrodef{SP}[SP]{Separation Property}
\acrodef{SRAM}[SRAM]{Static Random Access Memory}
\acrodef{STDP}[STDP]{Spike-Timing Dependent Plasticity}
\acrodef{STD}[STD]{Short-Term Depression}
\acrodef{STP}[STP]{Short-Term Plasticity}
\acrodef{STT-MRAM}[STT-MRAM]{Spin-Transfer Torque Magnetic Random Access Memory}
\acrodef{STT}[STT]{Spin-Transfer Torque}
\acrodef{SW}[SW]{Software}
\acrodef{TCAM}[TCAM]{Ternary Content-Addressable Memory}
\acrodef{TFT}[TFT]{Thin Film Transistor}
\acrodef{TLE}[TLE]{Temporal Lobe Epilepsy}
\acrodef{USB}[USB]{Universal Serial Bus}
\acrodef{VHDL}[VHDL]{VHSIC Hardware Description Language}
\acrodef{VLSI}[VLSI]{Very Large Scale Integration}
\acrodef{VOR}[VOR]{Vestibulo-Ocular Reflex}
\acrodef{WCST}[WCST]{Wisconsin Card Sorting Test}
\acrodef{WTA}[WTA]{Winner-Take-All}
\acrodef{XML}[XML]{eXtensible Mark-up Language}
\acrodef{divmod3}[DIVMOD3]{divisibility of a number by three}
\acrodef{hWTA}[hWTA]{hard Winner-Take-All}
\acrodef{sWTA}[sWTA]{soft Winner-Take-All}
\acrodef{EEG}[EEG]{Electroencephalography}
\acrodef{iEEG}[iEEG]{Intracranial EEG}
\acrodef{ECoG}[ECoG]{Electrocorticography}
\acrodef{SoC}[SoC]{System-on-Chip}
\acrodef{ECG}[ECG]{Electrocardiogram}
\begin{document}
\title{An Adaptive Event-based Data Converter for Always-on Biomedical Applications at the Edge}
\DeclareRobustCommand*{\IEEEauthorrefmark}[1]{%
  \raisebox{0pt}[0pt][0pt]{\textsuperscript{\footnotesize #1}}%
}
\author{
    \IEEEauthorblockN{
    Mohammadali Sharifshazileh \IEEEauthorrefmark{1}\,\IEEEauthorrefmark{2},
    Giacomo Indiveri\IEEEauthorrefmark{1}\,\IEEEauthorrefmark{2}} 

    \IEEEauthorblockA{\IEEEauthorrefmark{1}Institute of Neuroinformatics, University of Zurich and ETH Zurich;
    }
    \IEEEauthorblockA{\IEEEauthorrefmark{2}Zentrum für Neurowissenschaften Zürich.
    }

}

\maketitle
\begin{abstract}
  Typical bio-signal processing front-ends are designed to maximize the quality of the recorded data, to allow faithful reproduction of the signal for monitoring and off-line processing.
  This leads to designs that have relatively large area and power consumption figures.
  However, wearable devices for always-on biomedical applications do not necessarily require to reproduce highly accurate recordings of bio-signals, provided their end-to-end classification or anomaly detection performance is not compromised.
  Within this context, we propose an adaptive Asynchronous Delta Modulator (ADM) circuit designed to encode signals with an event-based representation optimally suited for low-power on-line spiking neural network processors.
  The novel aspect of this work is the adaptive thresholding feature of the ADM, which allows the circuit to modulate and minimize the rate of events produced with the amplitude and noise characteristics of the signal.
  We describe the circuit's basic mode of operation, we validate it with experimental results, and characterize the new circuits that endow it with its adaptive thresholding properties.
\end{abstract}


\section{Introduction}
\label{sec:intro}
Electrophysiological recordings are the basis for a large number of diagnostic tools in medicine~\cite{Peeples19}.
Most common are recordings of cardiac function, muscle function, nerve transmission, and brain function.
To achieve high reliability in off-line data analysis, or to provide medical experts with accurate bio-signal recordings, classical bio-medical devices are designed to produce very high signal quality recordings.
This is directly related to the performance of signal acquisition and digitization, which is evaluated in terms of signal-to-noise ratio, linearity, resolution and sampling rate.
State-of-the-art biomedical recording systems employ the most advanced integrated analog and digital signal processing methods to address those challenges using data compression, hardware sharing and time interleaving~\cite{Chae_etal08,Hosseiniejad_etal14,Sodagar_etal07}.
For these systems, digital filtering and heavy post processing introduce extensive power/bandwidth overhead that often leads to large area and high power consumption.

However, bio-medical applications at the edge, i.e., applications that use wearable devices for always-on monitoring and on-chip classification of bio-signals, typically do not need such high signal-to-noise ratios in their front-end processing stages.
Rather than reproducing the recorded signal as faithfully as possible, the goal of these devices is to operate continuously throughout the day, and detect events of interest, such as  heart beat anomalies~\cite{Bauer_etal19,Corradi_etal19} or \acp{HFO}, which have been identified as relevant bio-markers of the epileptogenic zone in epilepsy patients~\cite{Fedele_etal17,Sharifshazileh_etal21}.
Therefore, for these use-cases, the most stringent requirements are low power and area. 
A promising approach to satisfy these requirements, is to employ \acp{SNN} implemented using analog/digital circuits with dynamics that are well matched to those of the signals they are designed to process~\cite{Chicca_etal14,Indiveri_Sandamirskaya19}.
As these neuromorphic processors expect spikes as the input, it is important to develop bio-signal acquisition front-ends that pre-process and convert the biomedical sensory signals into streams of events.
A typical front-end of this type would comprise a \ac{LNA}~\cite{Atzeni_etal22,Liu_etal22}, followed by one or more filtering stages, and then by an \ac{ADM}~\cite{Corradi_Indiveri15} and a low-power \ac{SNN} processing module~\cite{Moradi_etal18,Frenkel_Indiveri22,Frenkel_etal19}.
\begin{figure}
   \centering
   \includegraphics[width=0.4\textwidth]{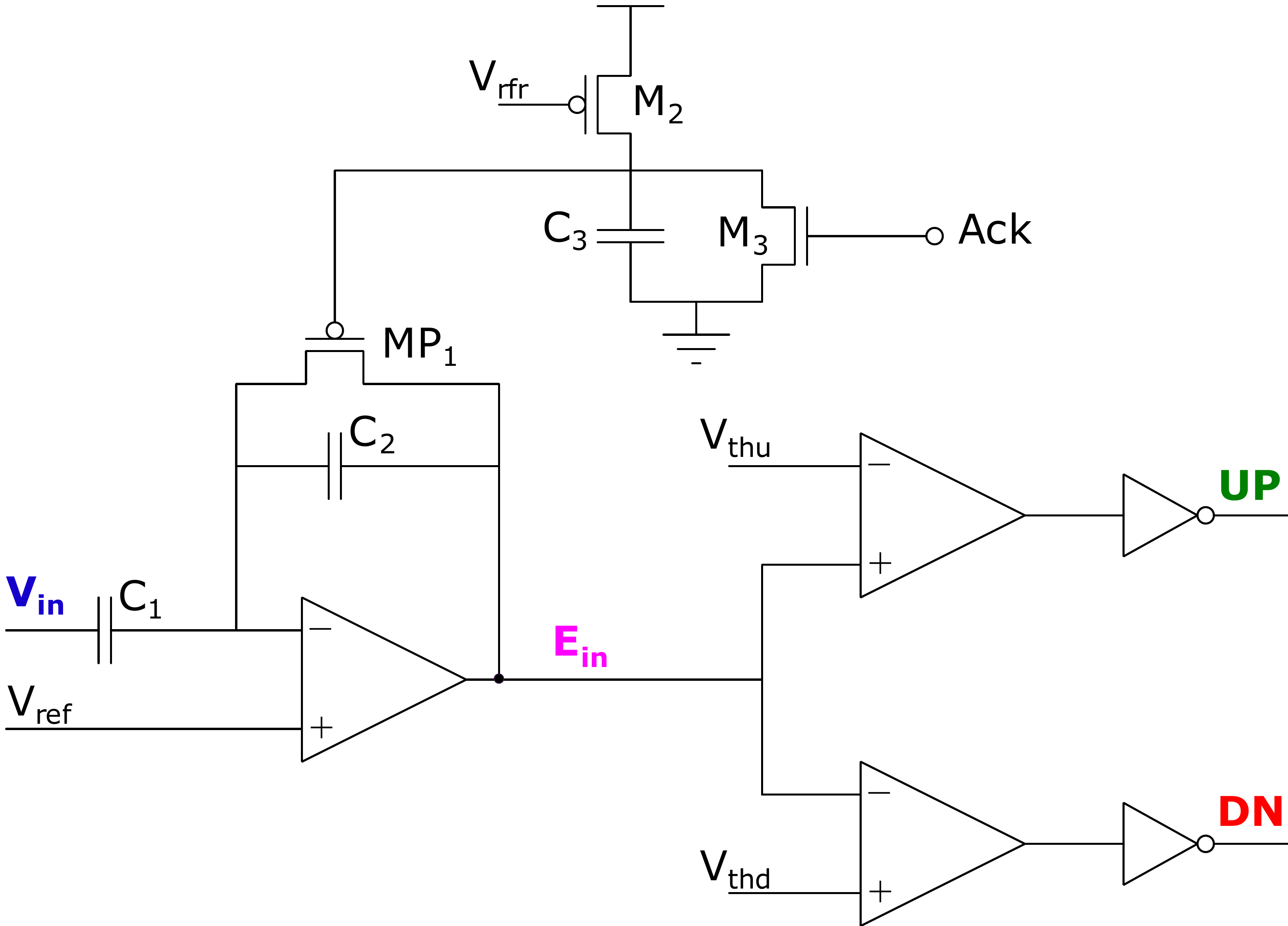}
   \caption{\ac{ADM} circuit. The analog input voltage $V_{in}$ is converted into UP and DN pulses (spikes) for further processing by \ac{SNN} circuits. The \ac{ADM} spike rate depends on the comparator thresholds $\mathtt{V_{thu}}$ and $\mathtt{V_{thd}}$.} 
   \label{fig:FAR}
\end{figure}
Several examples of applications that follow this approach have already been demonstrated using pre-recorded data-sets with fixed parameter settings~\cite{Corradi_Indiveri15,Bauer_etal19,Corradi_etal19,Ma_etal20a,Sharifshazileh_etal21}.
However, in real-world applications, the amplitude of the signals being measured may change over time (e.g., if the subject sweats, changes temperature, or if the wearable device shifts and moves), and the parameters of the system may need to be adjusted at run-time. 
In this work we focus on the \ac{ADM} block, analyzing how the circuit parameters affect the encoding of the signal, and we extend it by proposing a novel circuit that adjusts the \ac{ADM} parameters continuously, adapting to the changes in the average background activity, while preserving the sensitivity to large changes, or anomalies in the signal.
We propose analog circuits that that calculate on-line the best parameters and demonstrate the adaptive \ac{ADM} features on \ac{HFO} signals, for which it is important to produce events only for signals with amplitudes higher than the average background, and reduce, as much as possible, the generation of events for the background activity~\cite{Sharifshazileh_etal21}.


\section{The Asynchronous Delta Modulator (ADM)}
\label{sec:adm}

\begin{figure}
  \begin{subfigure}{0.475\textwidth}
    \includegraphics[width=\textwidth]{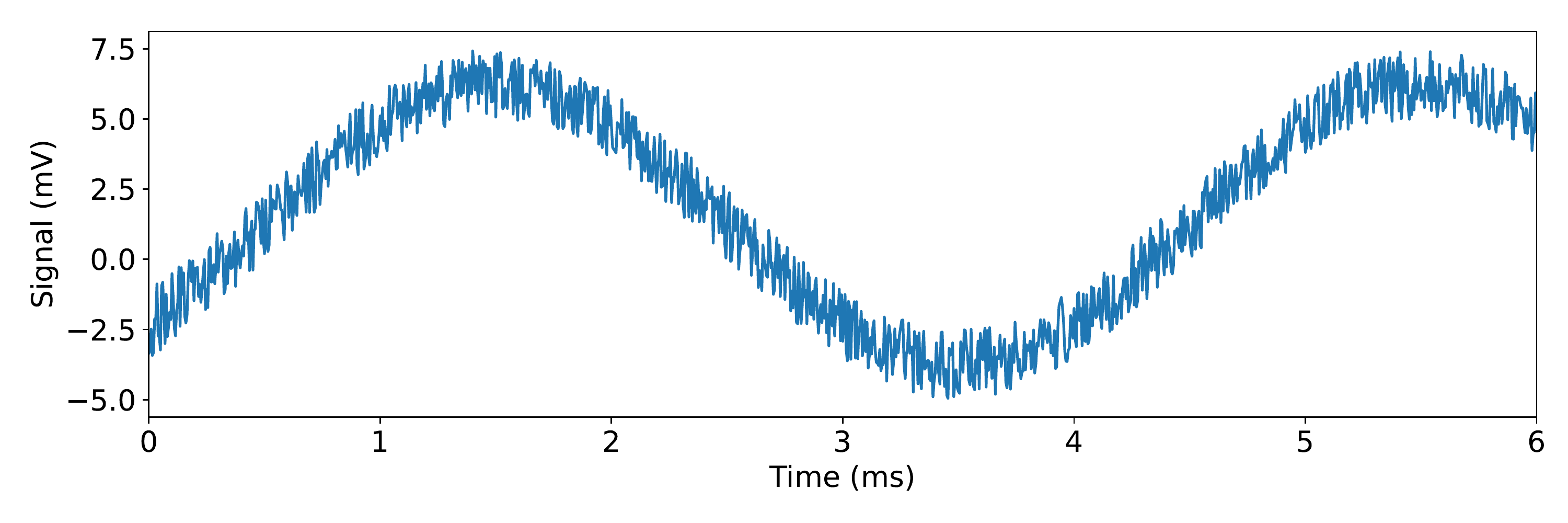}
    \subcaption{}
    \label{analog_inp_since}
  \end{subfigure}\\
  \begin{subfigure}{0.475\textwidth}
    \includegraphics[width=\textwidth]{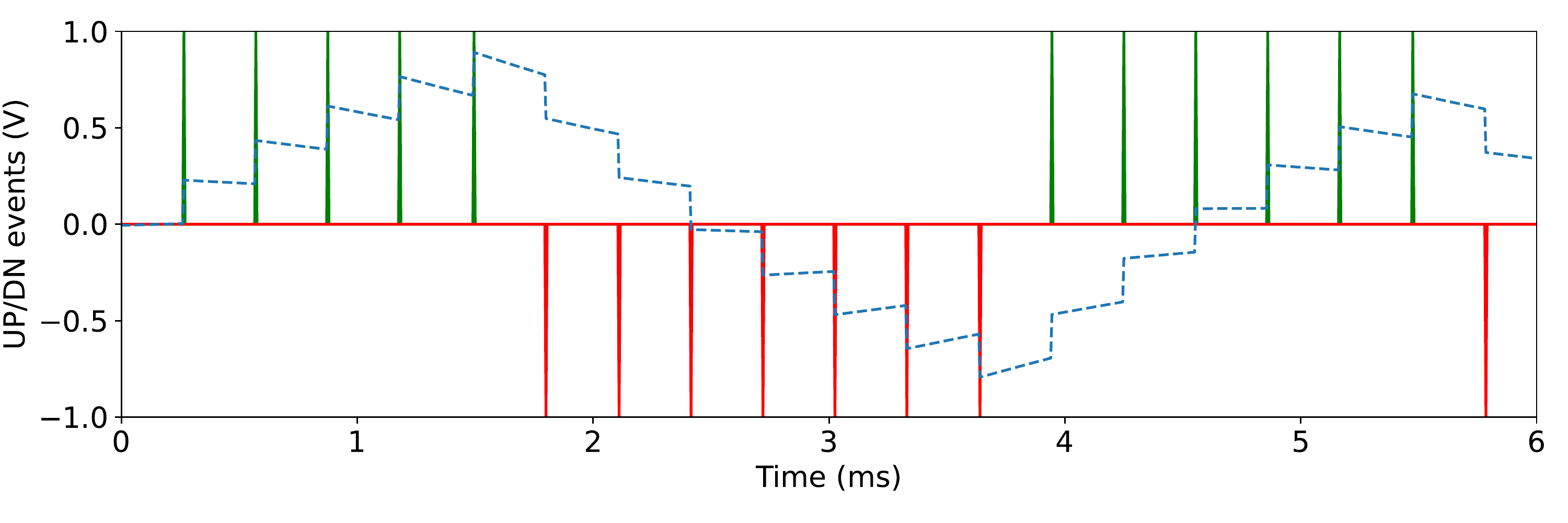}
    \subcaption{}
    \label{reconstriction_spikes}
  \end{subfigure}
  \caption{Experimental results. (a) Measured response of a \ac{LNA} circuit that receives in input a sine-wave attenuated to micro-Volt levels compatible with real bio-signals; (b) Response of the \ac{ADM} to the waveform of (a), producing the encoded \texttt{UP} and \texttt{DN} events. The dashed line represents the signal reconstructed in software using these events (with arbitrary units).}
   \label{fig:reconst}
\end{figure}

A circuit diagram of the \ac{ADM}, originally proposed in~\cite{Corradi_Indiveri15}, is shown in Fig.~\ref{fig:FAR}.
This circuit belongs to the family of delta modulation encoders first proposed in the 1960s for communication~\cite{Steele75} and recently adapted for event-based communication systems~\cite{Yang_etal14}.
It consists of an input \ac{OTA} with a capacitive divider gain stage, two comparators, and additional inverters that produce the \texttt{UP} or \texttt{DN} digital voltages when the change in the amplified version of the input signal, ($\mathtt{E_{in}}$), either increases above the reference voltage, $\mathtt{V_{ref}}$, by $\mathtt{V_{thu}}$ or decreases below $\mathtt{V_{ref}}$ by $\mathtt{V_{thd}}$.
When either of these digital signals switch, additional logic (not shown) asserts an active-high \texttt{Ack} signal, which in turn resets the \ac{OTA} output to $\mathtt{V_{ref}}$ and resets the \texttt{UP} and \texttt{DN} pulses, which are transmitted to further processing stages implemented using \ac{SNN} neuromorphic processors.
This reset state will be held for a ``refractory period''  determined by the values of the capacitor $\mathtt{C_3}$ and the bias voltage $\mathtt{V_{rfr}}$.
The number of the \texttt{UP} and \texttt{DN} pulses depends both on the features of the input signal (amplitude and frequency) and on the \ac{ADM} parameters: the comparator thresholds and the refractory period.

The maximum rate of pulses is hard-limited by the circuit's refractory period.
If we consider a sine-wave as the input signal, then the impact of signal related parameters and the circuit parameters on the rate of spikes generated by the \ac{ADM} can be expressed as: 

\begin{equation}
  R_{s}\propto \frac{A\cdot f}{T_{rfr}\cdot V_{th}}
  \label{eq:rate}
\end{equation}
where $R_{s}$ represents the \ac{ADM} spike rate for a sine wave of amplitude $\mathtt{A}$ and frequency $\mathtt{f}$; the parameter $\mathtt{T_{rfr}}$ represents the refractory period, and $\mathtt{V_{th} \vcentcolon =}$ $\mathtt{V_{thu}-V_{ref}=}$ $\mathtt{V_{ref}-V_{thd}}$ is the comparator threshold (assuming \texttt{UP} and \texttt{DN} thresholds are equal).

Figure~\ref{fig:reconst} shows experimental results measured from the \ac{ADM} fabricated using a standard 180\,nm \ac{CMOS} process \cite{Sharifshazileh_etal21}.
The input signal was a sine-wave that was attenuated down to a peak-to-peak range of 100\,$\mu$V, off-chip and passed through an \ac{LNA} with gain set to 38\,dB. The resulting input sine for the \ac{ADM} is depicted in Fig.~\ref{fig:reconst}a.
For this control experiment the threshold parameters for \texttt{UP} and \texttt{DN} spikes are kept constant and symmetric.
The events produced by the \ac{ADM} are plotted in Fig.~\ref{fig:reconst}b.

Accumulative addition/subtraction of the $\mathtt{V_{th}}$ value at the respective \texttt{UP} and \texttt{DN} spike times would result in a reconstruction of the input from the output spikes (e.g., see dashed line in Fig.~\ref{fig:reconst}b).
The reconstructed trace drifts away from the baseline due to the asynchronous and open-loop nature of delta-encoding, so to remove this drift effect, a high-pass filter is often applied to the staircase reconstructed signal.

\begin{figure}
   \centering
   \includegraphics[width=0.5\textwidth]{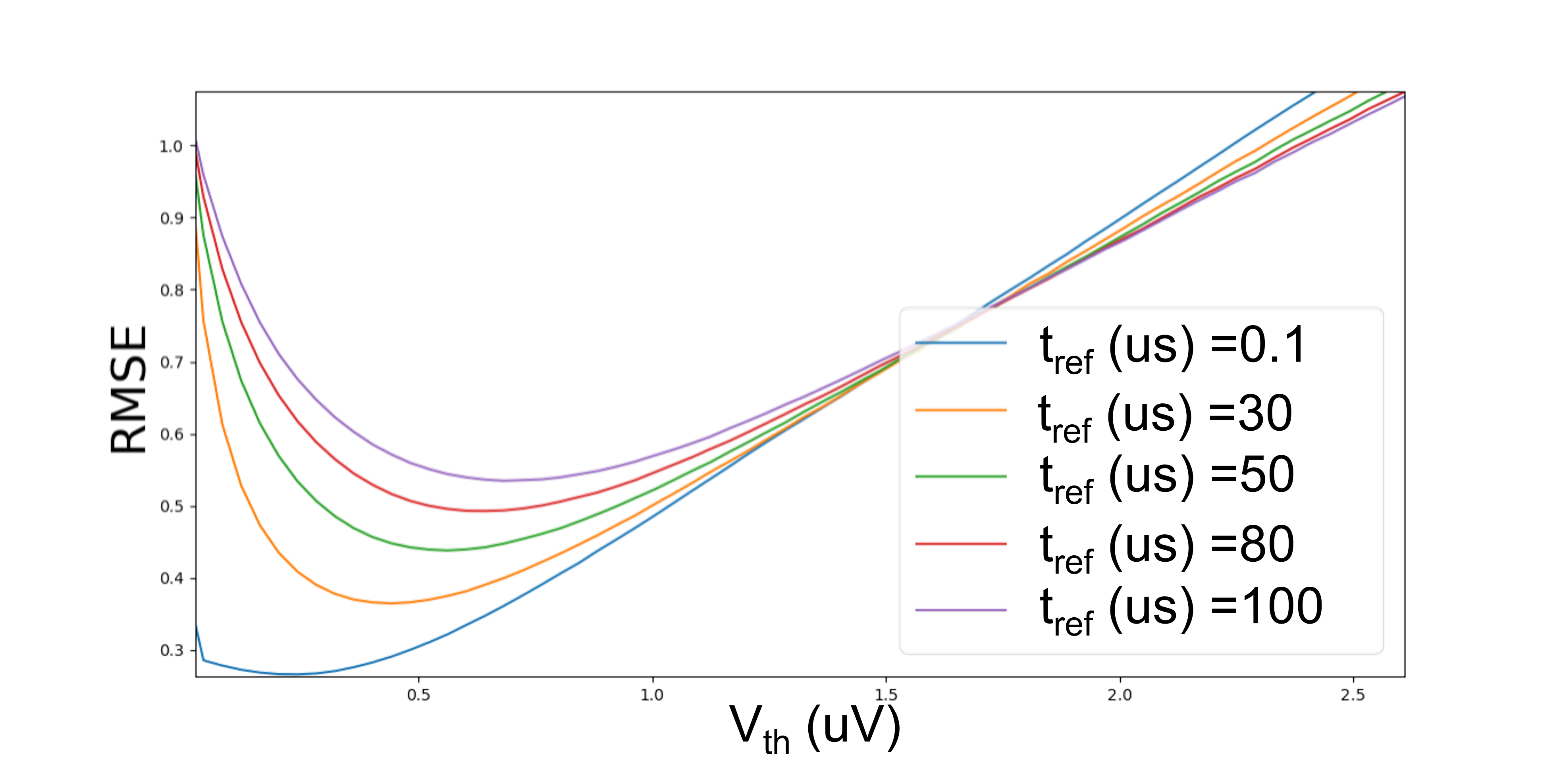}
   \caption{Root-mean-square of the \ac{ADM} spikes reconstruction error Vs. $V_{th}$ for various values of $t_{ref}$}
   \label{fig:RMSE}
\end{figure}

The accuracy of the data conversion can be quantified by calculating the \ac{RMSE} between the original signal and the reconstructed one. Figure~\ref{fig:RMSE} shows the effect of the two circuit parameters (refractory period and comparator threshold) on the reconstruction and the corresponding \ac{RMSE}. This analysis also shows that \ac{ADM} hyperparameters, $\mathtt{V_{th}}$ and $\mathtt{T_{rfr}}$, can be tuned to minimize the \ac{RMSE}. The fact that \ac{RMSE} has a non-zero minimum is due to the bounded spiking rate of the \ac{ADM}, limited by the refractory period.

\section{Adaptive ADM}
\label{sec:aadm}

\begin{figure}[t!]
  \begin{subfigure}{.5\textwidth}
    \centering
    \includegraphics[width=0.8\textwidth]{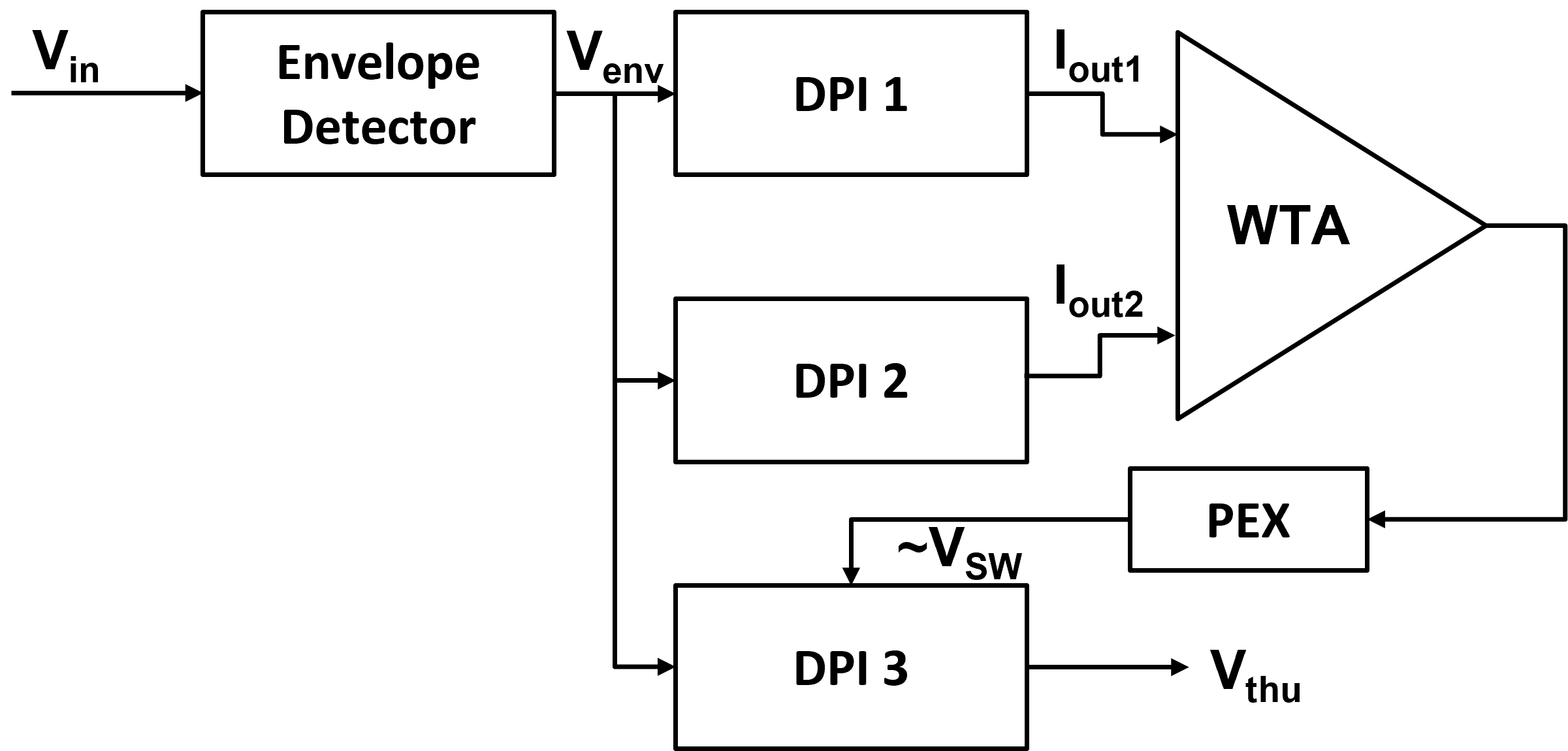}
    \label{block_diagram}
    \subcaption{}
  \end{subfigure} \\
  \begin{subfigure}[t]{.2\textwidth}
  \centering
    \includegraphics[width=0.8\textwidth]{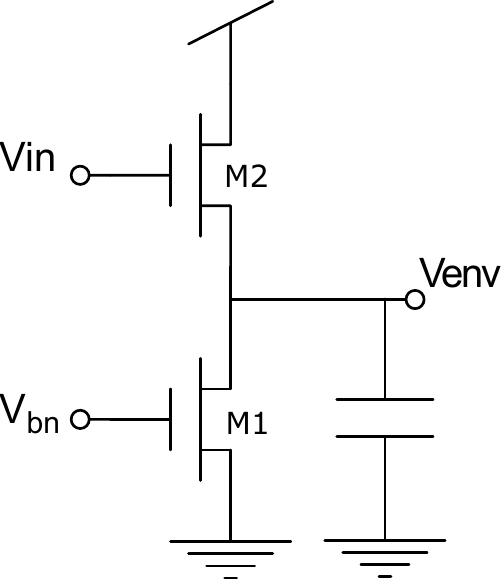}
    \label{sourcefollower}
    \subcaption{}
  \end{subfigure}\hfill
  \begin{subfigure}[t]{.3\textwidth}
  \centering
    \includegraphics[width=\textwidth]{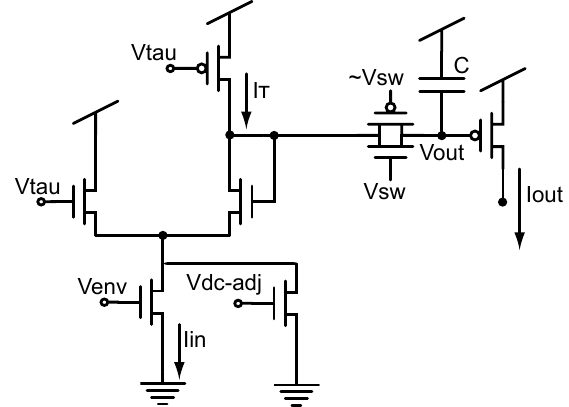}
    \label{dpi_switched}
    \subcaption{}
  \end{subfigure}
  \begin{subfigure}{.5\textwidth}
  \centering
    \includegraphics[width=0.55\textwidth]{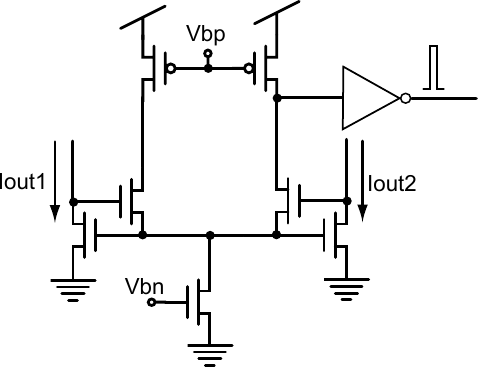}
    \label{winner_take_all}
    \subcaption{}
  \end{subfigure}
  \caption{Adaptive \ac{ADM} threshold-detection. (a) block diagram of the non-linear adaptive scheme adopted; (b) source follower integrator used to produce the signal representing the envelope of the input; (c) \ac{DPI} low-pass filter circuit with tunable cut-off frequency and controllable ``hold''
    switch. The DPI\,1 and DPI\,2 blocks do not have this gating switch, while the DPI\,3 block has it, controlled by the $\mathtt{V_{sw}}$ voltage; (d) \ac{WTA} circuit used to compare the output currents of DPI\,1 and DPI\,2 and produces an active high pulse when $\mathtt{I_{out1}<I_{out2}}$. This pulse gets extended to produce $\mathtt{V_{SW}}$.}
  \label{fig:circuits}
\end{figure}

To ensure autonomous long-time operation of bio-signal processing devices in real-world embedded applications, the \acp{ADM} should include variable gain and an automatic threshold adjusting mechanisms to compensate for changes that might occur in the signal (e.g., due to electrode movement or changing conditions of the subject). To this end, we have improved the \ac{ADM} in Fig.\ref{fig:FAR} in two directions.
First, we implemented a variable capacitor (2-bit switch, four capacitance values) in the feedback of the \ac{ADM} amplifier ($\mathtt{C_{2}}$ in Fig.~\ref{fig:FAR}) to add an extra degree of freedom on the gain of the front-end. This can help improving the dynamic range of bio-signal acquisition system, for capturing information in both high amplitude low-frequency signals and low-amplitude high-frequency ones.
The second and more important improvement comes in the form of the adaptive threshold detection module we present in Fig.~\ref{fig:circuits}.

The block diagram of this circuit is depicted in Fig.~\ref{fig:circuits}a. 
It operates by tracking the envelope of the signal through a source follower integrator (Fig.~\ref{fig:circuits}b); the envelope is then fed into two \acp{LPF} implemented using current-mode \ac{DPI} circuits~\cite{Bartolozzi_etal06} (see~Fig.\ref{fig:circuits}c), which are then used to produce a digital voltage to enable and disable a track-and-hold, depending on the changes in amplitude of the original signal. 

For sufficiently larger input signals, it has been shown that the output current of the \ac{DPI} circuit can be described with the following equation~\cite{Chicca_etal14} similar to a first-order low-pass filter:
\begin{equation}
\tau\frac{\mathrm{d} I_{out}}{\mathrm{d} t} + I_{out} = I_{in}
\end{equation}
where $\mathtt{\tau=\frac{CU_{T}}{\kappa I_{\tau }}}$

The \ac{DPI}1 and \ac{DPI}2 filters are configured with different time constant ($\tau$) parameters, such that their output currents $\mathtt{I_{out1}}$ and $\mathtt{I_{out2}}$ exhibit different step-response rise-times.
This differentiating behavior in the presence of high-amplitude-frequency activity, alongside with properly adjusted DC component of these currents, results in output traces that cross each other when sudden changes in the envelope of the input signal occur.

To compare the traces $\mathtt{I_{out1}}$ and $\mathtt{I_{out2}}$ and detect when they cross over we used a current-mode \acf{WTA} circuit, shown in~Fig.\ref{fig:circuits}d~\cite{Lazzaro_etal89}.
When the cross-over occurs the \ac{WTA} 
generates a pulse, which, for stability reasons, is extended using a pulse extender (PEX) circuit of the type presented in~\cite{Moradi_etal18}.
This extended pulse is used to open the switch of \ac{DPI}~3 and keep the circuit in hold-mode.
In this way the \ac{DPI}~3 capacitor only stores the values of the envelope computed on the average background activity of the signal, and ignores the large changes in the signal.
The result of this chain of operations is the setting of the \ac{ADM} spiking threshold parameter, $\mathtt{V_{th}}$, to a value proportional to the average background activity of the bio-signal for small fluctuations, and holding this value for periods of the signal that exhibit large fluctuations. This leads to an \ac{ADM} setting that produces a large number of spikes to faithfully encode signals of interests, and zero or few spikes in response to the noisy background activity. 

\begin{figure}
  \begin{subfigure}{.5\textwidth}
    \includegraphics[width=0.95\textwidth]{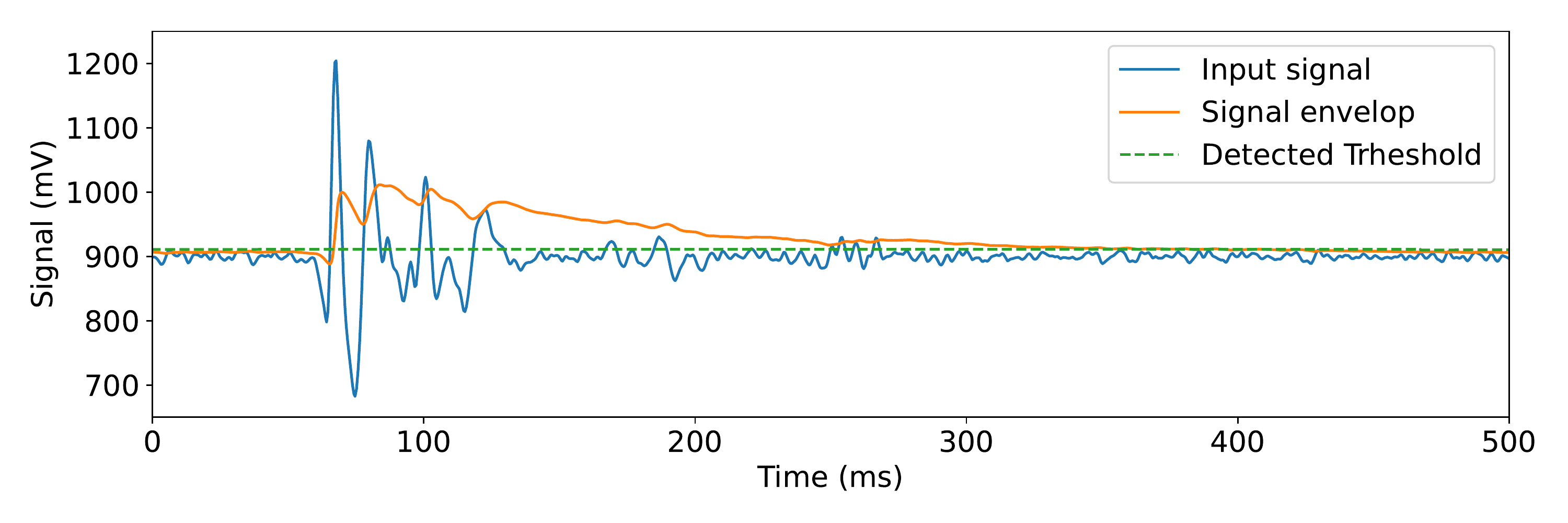}
    \label{traces-voltage}
  \end{subfigure} \\
  \begin{subfigure}{.5\textwidth}
    \includegraphics[width=\textwidth]{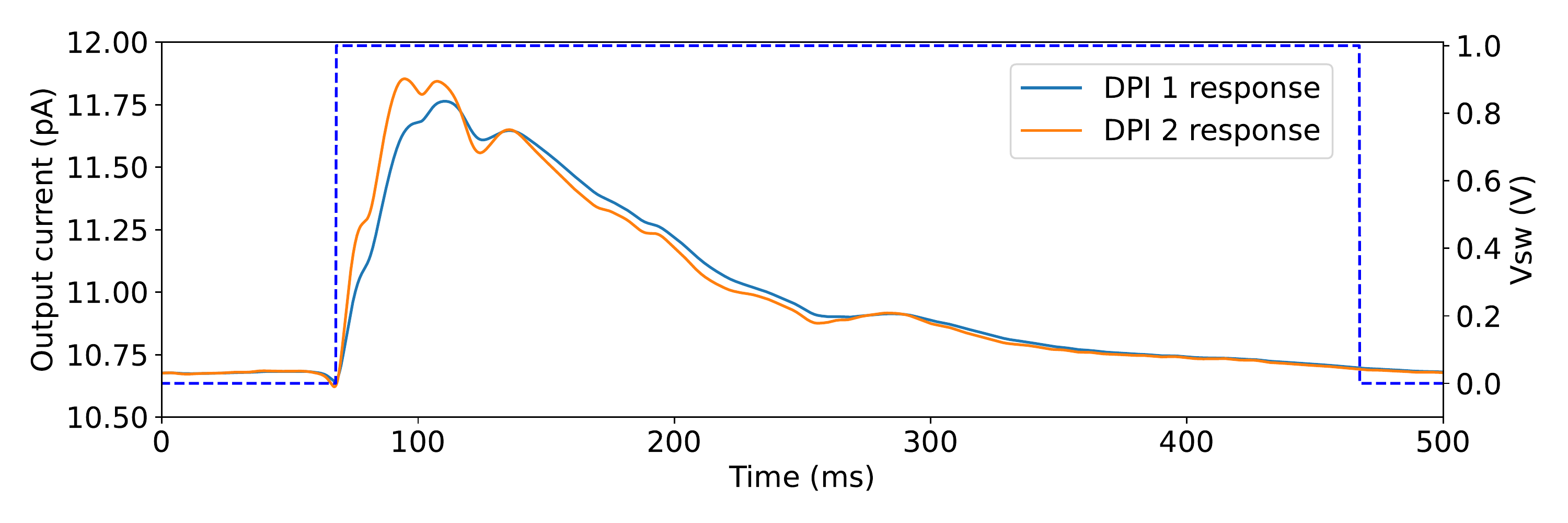}
    \label{traces-dpi}
  \end{subfigure}\\
  \begin{subfigure}{.5\textwidth}
    \includegraphics[width=0.95\textwidth]{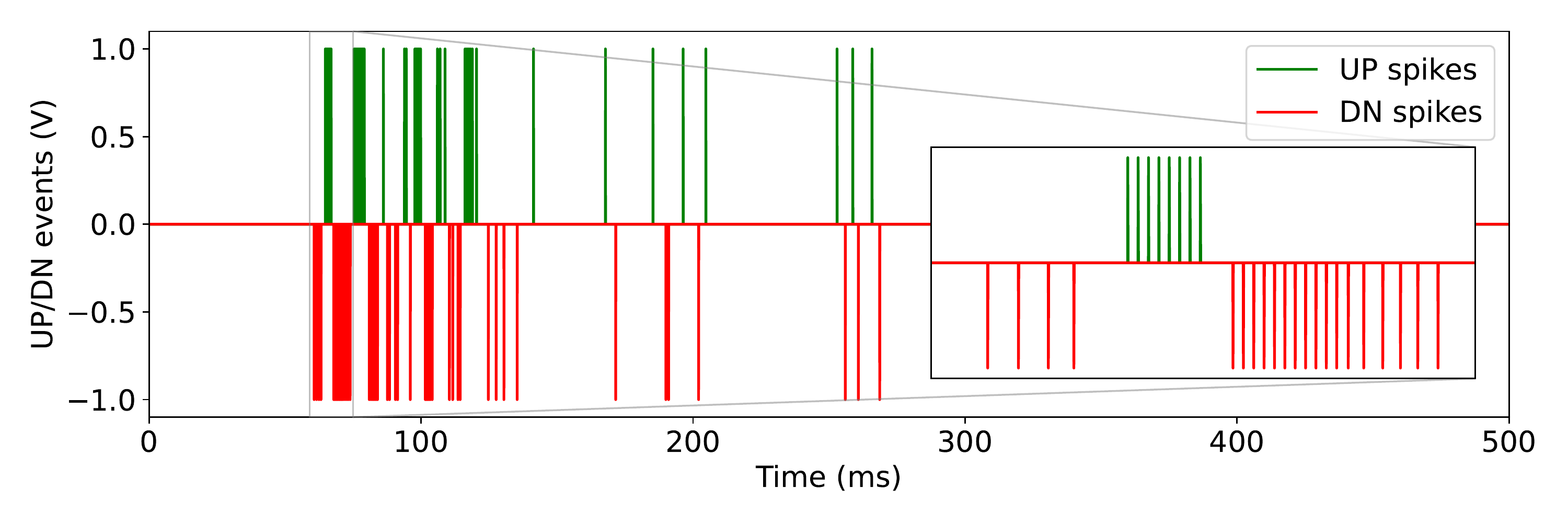}
    \label{traces-spikes}
  \end{subfigure}
  \caption{Adaptive \ac{ADM} simulation results. The top row shows a trace of pre-recorded \ac{iEEG}, it's envelope, and the $V_{th}$ parameter automatically detected. The middle row shows the output of the two \ac{DPI} filters used to detect outlier regions which do not belong to the average background activity. The dashed represents the digital voltage that gates the updates to the $V_{th}$ signal in these regions. The bottom row shows the UP and DN spikes generated by the \ac{ADM} circuit with the $V_{th}$ computed in this way. Note how the average background regions do not produce events. The inset in the plot shows the spikes produced for the first falling-rising-falling phases of the high-frequency oscillation.}
  \label{fig:plots}
\end{figure}

Circuit simulation results of the proposed adaptive threshold detection circuit are shown in Fig.~\ref{fig:plots}.
The input is a snippet of an \ac{iEEG} signal recorded with 2\,KHz sampling rate that is filtered in the 80-250\,Hz frequency band~\cite{Fedele_etal17b}.
This signal contains \acp{HFO}, characterized by high fluctuations from the average background activity.
The envelope of this signal, extracted by the source follower, is very stable during background activity and increases in response to the high-amplitude changes (see the orange trace in the top plot of Fig.~\ref{fig:plots}).
The output currents of the \acp{DPI}~1 and 2 in response to the envelope signal are depicted in the second row of Fig.~\ref{fig:plots}.
Note that the output current of \ac{DPI}~2 is always smaller than the \ac{DPI}~1 output current, when the envelope is constant.
However as the envelope increases rapidly, the \ac{DPI}~2 filter follows it more quickly than the \ac{DPI}~1 filter, and its output current crosses the \ac{DPI}~1 output current. This cross-over is detected by a \ac{WTA} circuit that continuously compares the two currents and produces a digital voltage pulse when the cross-over occurs.
This output pulse however is typically shorter than the full length of the event (it stays high only while $\mathtt{I_{out2}>I_{out1}}$).
So to ensure a stable $\mathtt{V_{th}}$, the pulse is extended to cover the full time-window of typical high-amplitude events, which depends on the nature of the  bio-signal, the time-constant of the envelope detector circuit, and time-constant of the \ac{DPI}~3 filter.
The result is a smooth and stable spiking threshold produced by the \ac{DPI}~3 circuit (see dashed green line in the top row of Fig.~\ref{fig:plots}).
The bottom row of Fig.~\ref{fig:plots} shows how the output spikes generated by the \ac{ADM} only encode the relevant events and ignore the background activity, thanks to the adaptive thresholding module.


\section{Discussion}
The major bottleneck holding back analog front-ends from being used with \ac{SNN} devices in low-power bio-signal processing edge applications is the data conversion block.
Optimizing the performance of conventional data converters is often performed by changing the design towards minimizing quantization error.
Indeed, knowing the characteristics of the input signal, one could optimize the hyperparameters of the \ac{ADM} to minimize the reconstruction error as shown in~Fig.\ref{fig:RMSE}.
However, as the goal of the \ac{SNN} processing pipeline is to classify or recognize features of interest in the input, rather than reconstructing the original signal faithfully, it is important to optimize resource usage (including area and power budget of the front-end) considering the end-to-end performance of the full system.
Mixed-signal \acp{SNN} heavily rely on analog front-ends that provide input spikes that encode  frequency- and amplitude-specific information present in analog input signal.
While most clinically relevant bio-signals tend to have sparse high-amplitude events, some such as \acp{LFP} are slow and gradual. So adding flexibility and adaptability to the analog front-ends is of critical importance.


\section{Conclusions}
\label{sec:conclusions}
In this article, we discussed about the benefits of using \ac{ADM} circuits in bio-signal acquisition front-ends. We demonstrated with experimental data their characteristics, and proposed a set of circuits for improving their performance, by endowing them with non-linear filtering properties that would allow them to change their parameters on-line, adapting to changes in their operating conditions or changes in the input signals.
We provided circuit simulation results, and validated them using clinically relevant bio-signals that contain long traces of average background activity, for which it is desirable to produce sparse spiking activity, if any, and short high frequency and high amplitude pulses, for which it is important to have \ac{ADM} parameters that can encode the signal with high accuracy.
Future developments will include integrating these circuits on a new design and testing them on \ac{EEG}, \ac{ECG}, and \ac{EMG} signals.


\cleardoublepage

\bibliographystyle{IEEEtran}
\bibliography{biblioncs.bib}

\end{document}